\documentstyle[twocolumn,aps,epsfig,floats]{revtex}

\begin{document}

\draft
\tolerance = 10000

\setcounter{topnumber}{1}
\renewcommand{\topfraction}{0.9}
\renewcommand{\textfraction}{0.1}
\renewcommand{\floatpagefraction}{0.9}

\twocolumn[\hsize\textwidth\columnwidth\hsize\csname
@twocolumnfalse\endcsname

\title{Statistical-thermodynamical foundations of anomalous diffusion}
\author{Dami\'an H. Zanette}
\address{\it Consejo  Nacional de Investigaciones  Cient\'{\i}ficas y
T\'ecnicas\\ Centro At\'omico Bariloche and Instituto Balseiro\\ 8400
Bariloche, R\'{\i}o Negro, Argentina}
\maketitle

\begin{abstract}
It is shown  that Tsallis' generalized  statistics provides a  natural
frame  for  the  statistical-thermodynamical  description of anomalous
diffusion. Within this generalized theory, a maximum-entropy formalism
makes  it  possible  to  derive  a  mathematical  formulation  for the
mechanisms that  underly L\'evy-like  superdiffusion, and  for solving
the nonlinear Fokker-Planck equation.
\end{abstract}

\vspace{1cm}

]

\section{Introduction: Diffusion processes}

Among  the  elementary  processes  that  underly  natural   phenomena,
diffusion is certainly one of the most ubiquitous.  In an ensemble  of
moving  elements  --atoms,  molecules,  chemicals, cells, or animals--
each element usually  performs, at a  mesoscopic description level,  a
random  path  with  sudden  changes  of  direction and velocity.  As a
result  of  this   highly  irregular  individual   motion,  which   is
microscopically driven  by the  interaction of  the elements  with the
medium, and of the elements with each other, the ensemble spreads out.
At a macroscopic level, this collective behavior is --in contrast with
the  individual  microscopic  motion--  extremely regular, and follows
very well defined, deterministic dynamical laws. It is precisely  this
smooth  macroscopic  spreading  of  an  ensemble  of  randomly  moving
elements that we associate with diffusion.

One of the first systematic observations of diffusion was made  by the
botanist  Robert  Brown  in  1828.  He  noticed  that pollen particles
dispersed in water exhibit a very irregular, swarming motion. In 1905,
Einstein  conjectured  that  this  ``Brownian  motion''  is due to the
interaction of pollen  with the water  molecules and, in  fact, proved
that microscopic particles suspended  in a liquid ``perform  movements
of such magnitude that they can be easily observed in a microscope, on
account of the molecular  motions of heat'' \cite{Einst}.  Since then,
Brownian motion is used as a synonym of diffusion.

A very suitable and very useful mathematical model for Brownian motion
is provided  by random  walks \cite{MW}.  In its  simplest version,  a
random walker is  a point particle  that moves on  a line at  discrete
time steps $\Delta t$. At each step, the walker chooses to jump to the
left or to the  right with equal probability,  and then moves a  fixed
distance $x$.   This stochastic  process can  be readily  generalized,
firstly, by allowing the walker  to move in a many-dimensional  space.
In  addition,  time  can  be  made  continuous by associating a random
duration with each jump or by introducing random waiting times between
jumps.  Finally, the length of each jump can be also chosen at  random
from a continuous set with a prescribed probability distribution.

Being  a  stochastic  process,  a  random  walk admits a probabilistic
description in  terms of  probability distributions  for the  relevant
quantities \cite{VK}.   In particular, one  is interested at  studying
the probability of finding the walker in a certain neighborhood $d{\bf
r}$ of point ${\bf r}$  --in general, in a $d$-dimensional  space-- at
time  $t$,  $P({\bf  r},t)\  d{\bf   r}$.   Note  that,  besides   its
interpretation as  a probability  distribution, $P({\bf  r},t)$ can be
related  to  the  density  in  an ensemble of noninteracting identical
random  walkers.   In  fact,  if  the  ensemble  contains $N$ walkers,
$n({\bf  r},t)=  N  P({\bf  r},t)$  stands  for  the  space density of
walkers.

Suppose that the random walk is defined in continuous time and  space,
with a  waiting time  probability distribution  $\psi(\tau)$ and  such
that the probability that the walker jumps from any point ${\bf r}$ to
${\bf r}+{\bf  x}$ is  $p({\bf x})\  d{\bf x}$.   The normalization of
probabilities imposes
\begin{equation}    \label{norm}
\int_0^\infty \psi(\tau) \ d\tau =1, \ \ \ \ \ \
\int p({\bf x})\  d{\bf x}=1.
\end{equation}
If, moreover, $\psi(\tau)$ and $p({\bf x})$ satisfy
\begin{equation}    \label{moms}
\langle \tau \rangle=\int_0^\infty \tau \ \psi(\tau) \ d\tau <\infty,
\ \ \ \ \langle x^2 \rangle= \int x^2 p({\bf x})\ d{\bf
x}<\infty,
\end{equation}
($x\equiv |{\bf x}|$) it can be proven that $P({\bf r},t)$ obeys the
diffusion equation \cite{MW}
\begin{equation}
\label{diffusion}
\frac{\partial P}{\partial t}=D \nabla^2_{\bf r} P,
\end{equation}
where $D \propto  \langle x^2 \rangle  / \langle \tau  \rangle$ is the
diffusion constant, or diffusivity. This equation has to be solved for
a  given  initial  condition  $P({\bf  r},0)$  with  suitable boundary
constraints.  The   density  $n({\bf   r},t)$  of   an  ensemble    of
noninteracting diffusing particles obeys the same equation.

A  typical  solution  to  the  diffusion  equation describes a density
profile that, as time elapses, is smoothed out and broadens. In  fact,
it can be straightforwardly shown  from the general solution that  the
width of the spatial distribution grows with time as
\begin{equation}    \label{width}
\langle r^2 \rangle = \int  r^2 P({\bf r},t)\ d{\bf r}
= 2dD t,
\end{equation}
where $d$ is  the space dimension.   Correspondingly, it can  be shown
that the  mean square  distance between  the present  position and the
initial position of a random  walk that satisfies Eq. (\ref{moms})  is
proportional  to  time.   This  proportionality  between  mean  square
displacement and time  is the fingerprint  of diffusion, as  it can be
used  experimentally,  numerically,  and  theoretically to detect this
kind of transport mechanism in a given natural process.

Though,  being  a  form  of  transport,  diffusion  is  inherently   a
nonequilibrium  process,  the  large-time  asymptotic  dynamics  of an
ensemble  of  diffusing  particles  can  be  described in the frame of
equilibrium statistical mechanics.  In fact, it is expected that,  for
very  large  times,  the  system  reaches  a  state of thermodynamical
equilibrium  with  the  medium  --and  between  the particles, if they
interact.   In  such  state,  the  diffusing  particles and the medium
participate of a  balanced interchange of  momentum and energy,  which
mantains the particles in their characteristic irregular motion.  Once
this situation is  reached, a connection  between the parameters  that
characterize thermodynamical equilibrium and particle dynamics  should
exist.  Einstein investigated  this problem in 1905  \cite{Einst}, and
concluded that diffusivity and temperature are proportional:
\begin{equation}    \label{Einstein}
D= \mu k_BT.
\end{equation}
Here $k_B$  is Boltzmann  constant, and  $\mu$ is  the mobility.   The
mobility is defined as the inverse of the friction coefficent, in  the
present case, of  the diffusing particles  in the medium  \cite{Kubo}.
The  Einstein  relation,  Eq.   (\ref{Einstein}),  provides  thus  the
expected connection between diffusion and thermodynamical equilibrium.

Despite  the  omnipresence  of  diffusion  as a transport mechanism in
natural processes, a different kind of transport underlies a  selected
--but ever growing-- class of systems. Due to various motivations most
of these systems  have recently attracted  very much attention.   They
range from turbulent fluids, to chaotic dynamical systems, to  genetic
codes (see next section).   In these systems, anomalous diffusion  --a
mechanism  closely  related  to   normal  diffusion,  but  with   some
qualitatively  different  properties--   drives  transport   processes
\cite{anom}.  Over the last few  years, it became more and more  clear
that  anomalous  diffusion  can  be  made  naturally  compatible  with
equilibrium  thermodynamics  if  the  Boltzmann-Gibbs  formulation  of
thermodynamics   is   replaced   by   Tsallis'.    This  compatibility
generalizes then the connection between normal diffusion and the usual
formulation  of  thermodynamics.   The  main  aim  of this paper is to
review  this  generalization,  commenting  on  some additional related
topics brought to light in recent work.

\section{Anomalous diffusion and L\'evy flights} \label{sect2}

Any transport mechanism which,  like diffusion, behaves at  mesoscopic
level  as  an  isotropic  random  process,  but  which  violates   Eq.
(\ref{width}), is generally refered  to as anomalous diffusion.   More
specifically, most of the  literature on anomalous diffusion  has been
devoted to processes where  the mean square displacement  $\langle r^2
\rangle$ varies with time as
\begin{equation}    \label{w}
\langle r^2 \rangle \propto t^{2/z},
\end{equation}
where $z$ ($\neq 2$) is  the dynamic exponent, or random  walk fractal
dimension, of the transport  process. Normal diffusion corresponds  to
$z=2$. For $z>2$, the growth  rate of the mean square  displacement is
smaller than in normal  diffusion, and transport is  consequently said
to  be  subdiffusive.  On  the  other  hand, for $z<2$ the mean square
displacement   grows   relatively   faster   and   transport  is  thus
superdiffusive. In the following, the attention will be mainly focused
on this latter case.

\subsection{Anomalous diffusion in Nature} \label{sect2.1}

As  advanced  above,  anomalous  diffusion  occurs  in a wide class of
natural systems and processes. In the realm of physics, a paradigmatic
example is given by  particle transport in disordered  media. Consider
the motion of particles in a medium containing impurities, defects, or
some  kind  of  intrinsic  disorder,  such  as in amorphous materials.
Examples are disordered lattices, porous media, and dopped  conductors
and semiconductors. In  these heterogeneous substrates,  particles are
driven by highly irregular forces, which determine a complex variation
of the local  transport coefficients. This  heterogeneity can in  fact
induce anomalous diffusion. For  instance, it has been  experimentally
shown  that   in  quasi-one-dimensional   ionic  conductors   such  as
hollandite (K$_{1.54}$Mg$_{0.77}$Ti$_{7.23}$O$_{16}$), where transport
is  very  sensitive  to  the  presence  of  impurities,  the dynamic
exponent is given by
\begin{equation}    \label{holl}
z \approx 1+ \frac{1}{\theta},
\end{equation}
with $\theta$ proportional to the temperature \cite{ber}.

A reasonable model for transport in heterogeneous media is provided by
a random walk  in a lattice  with quenched disorder  \cite{anom}. This
disorder applies to the depth  of the potential wells at  each lattice
site, and  to the  potential barriers  between sites.   Randomness  in
these parameters induces a distribution  for the time that the  walker
spends  at  each  site  before  hoping  to a neighbor. Generally, this
waiting time distribution,  $\psi(\tau)$, behaves as  $\psi(\tau) \sim
\tau^{-1-\mu}$ for $\tau\to\infty$.   For $\mu>1$, the first  relation
in Eq.  (\ref{moms}) holds,  and normal diffusion is observed.  On the
other hand, for $\mu <1$ the mean waiting time diverges and  diffusion
is anomalous.  The corresponding dynamic exponent for $0<\mu<1$ is
\cite{mac}
\begin{equation}    \label{heter}
z=2/\mu \ \ \ (d>2), \ \ \ \ \ z=2-d+d/\mu \ \ \ (d<2).
\end{equation}

A most important instance of anomalous diffusion in physics occurs  in
turbulent  flows.   In  fully  developed  turbulence,  fluid particles
exhibit very  irregular motion  over a  wide range  of space  and time
scales.  Based  on empirical motivations,  L.  Richardson  proposed in
1926 that the probability $P(R,t)$ that two fluid particles  initially
close to  one another  have a  separation $R$  at time  $t$ obeys  the
equation \cite{Richard}
\begin{equation}    \label{Rich}
\frac{\partial P}{\partial t}=\frac{\partial}{\partial R}
\left[ D(R) \frac{\partial P}{\partial R} \right].
\end{equation}
Comparing  with  (\ref{diffusion}),  it  is  clear that the Richardson
equation is a diffusion equation with space-dependent diffusivity. Its
solution  immediately  implies  $\langle  R^2  \rangle  \propto  t^3$,
indicating that the  relative motion of  particles in fully  developed
turbulence corresponds to anomalous diffusion with a dynamic  exponent
$z=2/3$, which is well into the superdiffusive regime.

Richardson's law has to be modified to take into account the fact that
the vorticity field in a turbulent flow is intermittent \cite{Mandel}.
This means that vorticity --and, in particular, turbulent activity and
dissipation--  is  concentrated  on  a  relatively small volume in the
whole system, which happens to be a fractal set. Experiments \cite{df}
suggest that the fractal dimension  of this set is $d_f=2.8\pm  0.05$.
Incorporating  this  correction,  it  can  be shown \cite{turbul} that
$\langle R^2 \rangle \propto t^{12/(1+d_f)}$, or
\begin{equation}    \label{zturb}
z=\frac{1+d_f}{6} \approx 0.63.
\end{equation}

A somewhat more abstract form of anomalous diffusion is present in the
evolution of  chaotic Hamiltonian  dynamical systems  in phase  space.
Hamiltonian systems are characterized by volume conservation in  phase
space, as stated  by the Liouville  theorem. The domain  occupied by a
given  set  of  initial  conditions  in  phase  space  can be strongly
distorted  under  the  effect  of  evolution,  but  its volume remains
constant. Mechanical processes that  preserve energy are instances  of
Hamiltonian systems, but a huge host of systems --both continuous  and
discrete in  time-- is  known to  belong to  the same  class. Since  a
single  Hamiltonian  system  can  exhibit  both  regular  and  chaotic
evolution  by  simply  changing  the  initial condition, the dynamical
geometry of its phase space is usually extremely intrincate. Zones  of
nested regular trajectories which  alternate with chaotic regions  are
typically found at many scales, displaying selfsimilar structures.  In
the bulk of chaotic  regions trajectories are extremely  irregular and
resemble  random  paths.   On  the  other  hand,  when approaching the
boundary with a  regular region, the  same trajectory can  temporarily
become much simpler and smoother. Consequently, as it evolves in phase
space  along  a  chaotic   orbit,  a  Hamiltonian  system   alternates
intermittently between zones of  highly complex behavior and  a regime
of almost regular dynamics.   Globally, this motion can be  thought of
as a  stochastic process,  and turns out to have the same  statistical
properties as anomalous diffusion.

A case  studied in  detail in  the literature  is the so-called Q-flow
\cite{Qflow}. It is defined as a three dimensional velocity field
\begin{equation}    \label{Qf}
\begin{array}{ll}
\dot x = & \partial\Psi / \partial y +\epsilon \sin z \\
\dot y = & \partial\Psi / \partial x +\epsilon \cos z \\
\dot z = & \Psi
\end{array}
\end{equation}
with
\begin{equation}    \label{Psi}
\Psi (x,y) = \sum_{j=1}^k \cos \left[ x \cos (2\pi j/k)+y \sin
(2\pi j/k) \right].
\end{equation}
Here $\epsilon$ is a parameter  and $k$ is an integer  that determines
the symmetry of the flow. The solution to Eqs. (\ref{Qf}) is a complex
trajectory that wanders  in an infinite connected net of  channels  of
width of order  $\epsilon$, inside which  the trajectory looks  like a
random contour \cite{Nat}.  Numerical measurements of the  statistical
properties of these diffusion-like trajectories show that the  dynamic
exponent  $z$  to  be  associated  with  them fluctuates strongly as a
function of $\epsilon$ \cite{Qflow}.  For $k=6$ and $0.8 < \epsilon  <
1.8$, $z$ varies in the interval
\begin{equation}    \label{zQf}
1<z<2,
\end{equation}
making apparent that the motion is supperdiffusive, as in  turbulence.
Anomalous   diffusion   has   also   been   observed   in  dissipative
(non-Hamiltonian)  dynamical  systems,  both  in  simulations  and  in
experiments. For example, a dynamic exponent $z \approx 1.2$ has  been
measured in the Taylor-Couette flow\cite{TC}.

As stated in the  Introduction, anomalous diffusion is  not restricted
to physical systems. This kind of transport has in fact been  detected
to underly several biological  processes. Some sectors of  genomic DNA
sequences, for instance, are  known to exhibit statistical  properties
analogous  to  anomalous  diffusion.   To  stress this correspondence,
``DNA walks'' have been  defined \cite{DNA}. DNA, which  codes genetic
information,  is  a  large  molecule  in  the  form  of  a  chain   of
nucleotides. Each nucleotide contains either a purine or a  pyramidine
base. Two  purines --the  adenine (A)  and the  guanine (G)--  and two
pyramidines  --the  cytosine  (C)  and  the  thymine (T)-- are in turn
present in the DNA chain.  Therefore, the information code in DNA is a
symbolic chain of four letters:  A, C, G and T.  Amazingly, within DNA
only a small portion does  code information for protein building  (3\%
in the human genome) whereas other zones are noncoding, their specific
role being  unknown.   A one-dimensional  DNA walk  is constructed  by
sequentially running over the chain of nucleotids. Each time a  purine
is found  the walker  jumps rightwards,  whereas when  a pyramidine is
found the jump occurs leftwards. For instance,
$$
\cdots \mbox{ACGCTGAGTG} \cdots \ \to  \
\cdots +-+--+++-+ \cdots
$$
where $+$ stands for  jumps to the right  and $-$ stands for  jumps to
the  left.   In  this  DNA  walk,  systematic  deviations  from normal
diffusive  behavior  derive  from   long-range  correlations  in   the
nucleotide sequence. It has  been found that the  DNA walk is in  fact
statistically  identical  to  normal  diffusion  in  the  zones of the
genomic chain that code information.  On the other hand, in  noncoding
sequences the DNA walk is  analogous to superdiffusion.  In  the human
beta-globin  chromosomal  region  the  dynamic exponent is $z\approx
1.4$ \cite{DNA1}.

A less involved instance of anomalous diffusion in biology appears  in
the flight patterns of certain birds. In particular, it has been found
\cite{albat} that in the foraging behavior of the wandering  albatross
({\it Diomedea exulans}) the flight-time intervals exhibit a power-law
distribution. This results in a anomalous diffusive-like motion which,
according to field measurements on the Bird Island, South Georgia, has
a  dynamic  exponent  $z  \approx  1.2$.   The kind of flight patterns
observed in these seabirds is supposed to reflect a complex  structure
in the underlying ecosystem,  especially, in the spatial  distribution
of the exploited environment.  It has been suggested that  albatrosses
specialize in long journeys of random foraging, searching for patchily
and  unpredictably   dispersed  prey   over  several   million  square
kilometers. As discussed in  the next section, anomalous  diffusion is
inherently  related  with  fractal  geometry,  scale-invariance,   and
self-similarity,  which  in  the  present  example  seem  to drive the
predator-prey dynamics.

Of course, the previous collection of examples of anomalous  diffusion
in Nature is not at all  exhaustive, but only pretends to give  a hint
on the variety of systems driven by this kind of transport. For a more
detailed account, the reader is refered to the review by Bouchaud  and
Georges \cite{anom}. This review is also an excellent reference to the
mathematical  treatment  of  anomalous  diffusion,  which  is  briefly
introduced in the following.

\subsection{Random-walk models of anomalous diffusion} \label{sect2.2}

In view of the efficacy  of random walks in modeling  normal diffusion
at a mesoscopic  level, it is  desirable to find  a similar stochastic
model describing anomalous diffusion.   It has already been  mentioned
in Section \ref{sect2.1} that introducing a waiting time  distribution
which,  for  large  waiting  times,  behaves  as  $\psi  (\tau)   \sim
\tau^{-1-\mu}$ with $\mu<1$,  produces the anomalous  dynamic exponent
given  in  Eq.   (\ref{heter}).   In  this  case $z>2$ and, therefore,
transport is subdiffusive.  Note  that the source of anomaly  in these
random walks is  the divergency of  the average waiting  time $\langle
\tau\rangle$.  For $\mu<1$, the long-tailed distribution  $\psi(\tau)$
allows for very long  waiting times with relatively  high probability,
violating thus the  first relation in  Eq.  (\ref{moms}).   These long
waiting  times  produce  an  overall  reduction  of  efficiency in the
transport mechanism  with respect  to normal  diffusion, and  leads to
subdiffusive behavior.

Taking into account the previous argument, in can be expected that the
violation of the  second of relations  (\ref{moms}) will lead,  on the
other hand, to superdiffusion. In fact, having a divergent mean square
displacement  $\langle  x^2\rangle$  requires  the  jump   probability
distribution $p({\bf  x})$ to  have a  long tail  for large  $x$, in a
sense to be made  precise immediately.  This  long-tailed distribution
would  produce,  with  relatively  high  probability, very long jumps.
Globally, this transport mechanism  should result more efficient  than
normal diffusion, and superdiffusive behavior is expected.

It can be easily shown that, in $d$-dimensional space, the mean square
displacement diverges if
\begin{equation}    \label{asymp}
p({\bf x})\sim \frac{1}{x^{d+\gamma}}
\end{equation}
for large  $x$, with  $\gamma<2$.   Note that,  for such distribution,
normalization requires $\gamma>0$. It is therefore to be expected that
a random walk with a jump distribution $p({\bf x}) \sim x^{-d-\gamma}$
with $0<\gamma<2$ does  not model normal  diffusion, but some  kind of
superdiffusive motion.  Of course, a large score of functions  satisfy
Eq. (\ref{asymp}), and are thus candidates to play the role of a  jump
distribution  for  a  superdiffusive  random  walk. Among them, L\'evy
distributions have been studied in detail.

L\'evy  distributions  \cite{Lev} are  defined  through  their Fourier
transform, which reads
\begin{equation}    \label{Levy}
p({\bf k}) = \int \exp(i{\bf k}\cdot {\bf x})\
p({\bf x})\ d{\bf x} = \exp(-b k^\gamma),
\end{equation}
where $b$ is  a positive constant,  and $k\equiv |{\bf  k}|$. Although
the antitransform $p({\bf x})$ has no analytical expression, it can be
shown that it satisfies Eq. (\ref{asymp}). Moreover, if $\gamma<2$ the
positivity of $p({\bf x})$ is  insured. The relatively simple form  of
this jump distribution in the Fourier representation makes it an ideal
tool for analytical manipulation. However, the main interest of L\'evy
functions in the mathematical  theory of distributions comes  from the
fact that  they are  stable. Essentially,  this means  that two L\'evy
functions  with  the  same  L\'evy  exponent  $\gamma$  produce,  upon
convolution, a third L\'evy function with the same exponent. This  can
be readily proven in the Fourier representation, where the convolution
transforms into ordinary product:
\begin{eqnarray}
p_1({\bf k}) p_2({\bf k})&=&\exp(-b_1k^\gamma)\exp(-b_2k^\gamma)
\nonumber \\
&=&\exp[-(b_1+b_2)k^\gamma]= p_3({\bf k}).\label{conv}
\end{eqnarray}

L\'evy functions are not  the only stable distributions,  the Gaussian
$p({\bf x})\propto \exp(-x^2)$ being probably the best-known  example.
The  (one-dimensional)  Cauchy  distribution,   $p(x)  \propto  (1   +
x^2)^{-1}$,  is   another  instance.   Stable  distributions   play  a
fundamental  role  in  probability  theory  since  according  to   the
central  limit  theorem  --which  is  usually  stated for the Gaussian
function--  the  addition  of  random  variables  tends  to  one  such
distribution.  In particular, as  P.  L\'evy demonstrated through  his
generalization of the  Gaussian central limit  theorem \cite{Lev,CLT},
adding  random  variables   with  a  power-law   distribution  as   in
(\ref{asymp}) --whose second moment $\langle x^2 \rangle$ diverges for
$\gamma<2$-- leads asymptotically to a L\'evy distribution.

Another important property of L\'evy distributions, which is reflected
in its power-law large-$x$ asymptotic behavior, Eq. (\ref{asymp}),  is
the absence of characteristic length scales.  This implies that random
walks with L\'evy jump distributions have self-similar properties.  In
particular, it can  be shown that  the set of  points visited by  this
kind of random walk is a fractal of dimension $\gamma$ \cite{Proc}. As
a  consequence,  these  distributions  ubiquitous  in  the  realm   of
self-similarity geometry --the geometry of fractals.

\begin{figure}
\begin{center}
\psfig{figure=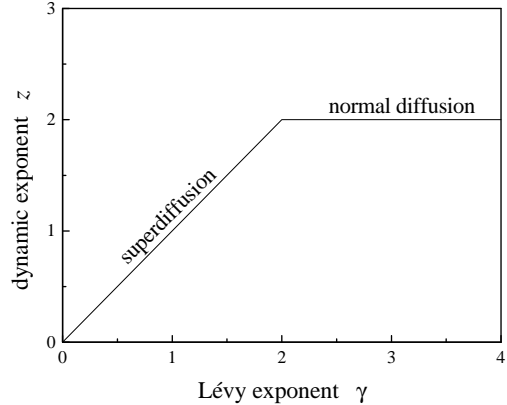,width=\columnwidth}
\end{center}
\caption{The  diffusion  dynamic  exponent  $z$  as  a function of the
L\'evy exponent $\gamma$.}
\label{f2}
\end{figure}

A discrete-time  random walk  whose jump  distribution is  given by  a
L\'evy function as in Eq.  (\ref{Levy}) is called a L\'evy  flight. It
has been  suggested \cite{Compte}  that the  Fourier transform $P({\bf
k},t)$ of  the probability  distribution of  finding the  walker at  a
given point at time $t$ satisfies the evolution equation
\begin{equation}    \label{dif}
\frac{\partial P}{\partial t}({\bf k},t) = - D_\gamma k^\gamma
P({\bf k},t).
\end{equation}
This  equation  generalizes,  in   the  Fourier  representation,   the
diffusion equation (\ref{diffusion}). A straightforward dimensionality
analysis  shows  that  the  anomalous  diffusivity is $D_\gamma\propto
b/\Delta t$, where  $\Delta t$ is  the time step  of the random  walk.
In free space, equation (\ref{dif}) can be readily solved:
\begin{equation}    \label{Pk}
P({\bf k},t)=P({\bf k},0) \exp(-D_\gamma k^\gamma t).
\end{equation}
For a delta-like initial  condition, $P({\bf r},0)=\delta ({\bf  r})$,
one has $P({\bf k},0)=1$ and, thus,
\begin{equation}    \label{Pr}
P({\bf r},t)=(2\pi)^{-d}\int \exp(-i{\bf k}\cdot {\bf r}-D_\gamma
k^\gamma t) \ d{\bf k}.
\end{equation}
This function remains unchanged, except for  a  constant  factor,   if
both  space  and  time are conveniently rescaled, $P(\alpha^{1/\gamma}
{\bf  r}, \alpha   t)=\alpha^{-d/\gamma} P({\bf  r},t)$.  Using   this
scale   invariance,  one  can  write
\begin{equation}    \label{scaling}
P({\bf r},t) = t^{-d/\gamma}\Pi (r/t^{1/\gamma}),
\end{equation}
where  $\Pi$ is  a   function of  a single variable
\cite{Foged}. In turn, this implies
\begin{equation}    \label{msd}
\langle r^2 \rangle \propto t^{2/\gamma},
\end{equation}
for $0<\gamma<2$.   This result,  which has  been here  derived for  a
delta-like  initial  distribution,  can   be  generalized  by   simple
superposition  to  more  general  initial  conditions. It shows that a
L\'evy  flight  with  $0<\gamma<2$  represents  superdiffusion  with a
dynamic exponent  $z=\gamma$. On  the other  hand, for  power-law jump
distributions  with  $\gamma>2$  the  dynamic  exponent corresponds to
normal diffusion, $z=2$ (Figure \ref{f2}).

The fact that in L\'evy flights the mean square displacement  $\langle
x^2 \rangle$ of a single step diverges, implies --in contrast with Eq.
(\ref{msd})-- that the mean square displacement of the walker after  a
certain time  is, on  the average  over infinitely  many realizations,
also  infinite.   The  arguments  used  to  derive Eq. (\ref{msd}) are
therefore of  limited validity  \cite{Compte,Foged},  and have  to be
taken {\it cum grano salis}. The result (\ref{msd}) is expected to  be
valid for finite times, i.e. in a certain portion of the whole  random
walk, and on averages over a finite number of trajectories.  The  same
result would be valid during  a certain time if the  jump distribution
$p({\bf x})$ is a L\'evy function  in some (large) range of values  of
$x$, but has a cutoff  for sufficently large $x$ \cite{Tsa}.  In spite
of this  drawback, L\'evy  flights provide  a very  powerful tool  for
modeling superdiffusion because of the mathematical properties of  the
L\'evy  distributions,  summarized  above.   They  are  thus  a   very
satisfactory starting point  as a model  for studying the  statistical
mechanics  of  superdiffusive  transport,  generalizing  the   results
outlined in the Introduction for normal diffusion.

\begin{figure}
\begin{center}
\psfig{figure=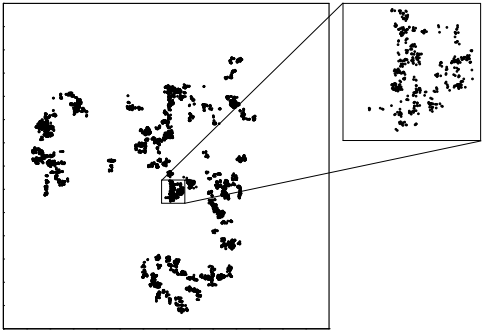,width=\columnwidth}
\end{center}
\caption{The first $10^4$ points  visited by a two-dimensional  random
walk  generated  by  a  power-law  jump  distribution  with   exponent
$\gamma=1.5$,  starting  at  the  center  of  the  main  frame.    The
amplification  illustrates   the  self-similar   properties  of   this
process.}
\label{f1}
\end{figure}

Before   passing   to   the   discussion   of   superdiffusion   in  a
statistical-mechanical frame, a comment  is in order on  the numerical
simulation  of  superdiffusive  random  walks.   Due to the cumbersome
properties of L\'evy  distributions in real  space \cite{Proc}, it  is
not convenient  --in numerical  calculations-- to  work directly  with
these  functions.   Rather,  power-law  distributions  with  the  same
asymptotic properties as  L\'evy's, Eq. (\ref{asymp}), are used.   For
instance, one can take
\begin{equation}    \label{num}
p({\bf x}) = \frac{{\cal N}}{(1+x)^{d+\gamma}},
\end{equation}
with ${\cal N}$  a normalization constant.   The Fourier transform  of
these distributions behaves precisely  like a L\'evy distribution  for
small $k$, $p({\bf k}) \approx 1-bk^\gamma$. Also, they show the  same
scale invariance for large $x$, wich leads to self-similar  properties
in the associated random walks.   In Figure \ref{f1} the first  $10^4$
points  visited  by  a  random  walk  in two dimensions, with the jump
distribution given in (\ref{num})  and $\gamma=1.5$, are shown.   Note
the clustered, fractal-like structure of this set of points.

\section{Maximum-entropy formalism for anomalous diffusion}

Entropy plays  a central  role in  the foundations  of equilibrium and
nonequilibrium statistical mechanics. It  is well known from  the work
by  L.   Boltzmann  and  others  that  entropy provides a natural link
between  nonequilibrium  processes  and  their  asymptotic  states  of
thermodynamical  equilibrium.   In  addition,  the  whole  theory   of
equilibrium statistical  mechanics can  be derived  from a variational
formalism for the entropy,  as follows.  Define  the entropy $S$ as  a
functional of the probability  distribution $p_i$ over the  states $i$
of a given system,
\begin{equation}    \label{Si}
S[p]  =-k_B \sum_i p_i \ln p_i,
\end{equation}
where $k_B$ is Boltzmann constant. Find then the values of $p_i$  that
maximize  $S[p]$,  taking  into  account the normalization constraint,
$\sum_i  p_i=1$,  and  --if  required  by particular conditions of the
system under study-- any additional constraint on $p_i$. The value  of
$p_i$ resulting from this maximization procedure gives the probability
of finding the  system in state  $i$ when thermodynamical  equilibrium
has been reached. For  instance, introducing the canonical  constraint
$\sum \epsilon_i p_i=E$, where $\epsilon_i$ is the energy of state $i$
and $E$  is the  thermodynamical energy,  the maximization  of entropy
produces   the   well-known   Boltzmann   distribution   $p_i  \propto
\exp(-\beta\epsilon_i)$ \cite{Kubo}.

\subsection{Traditional formalism}

As a  starting point  for including  normal diffusion  in the frame of
equilibrium   statistical   mechanics,   the   procedure   of  entropy
maximization  has  been  applied   to  obtain  the  jump   probability
distribution $p({\bf  x})$ in  a discrete-time  random walk \cite{MS}.
In this case, entropy  is defined as a  straightforward generalization
of (\ref{Si}),
\begin{equation}    \label{Sx}
S[p]  =-k_B \int p({\bf x}) \ln [\sigma^d p({\bf x})]\ d{\bf x}.
\end{equation}
Here $\sigma$ is  a characteristic length,  whose meaning will  become
clear immediately. The distribution  $p({\bf x})$, in fact,  has units
of length to the power $-d$. The maximization of $S[p]$ is carried out
taking into account the normalization of $p({\bf x})$,
\begin{equation}    \label{normal}
\int  p ({\bf x})\ d{\bf x} =1,
\end{equation}
and  imposing  the additional constraint
\begin{equation}    \label{const}
\int x^2 p ({\bf x})\ d{\bf x} =\sigma^2 d,
\end{equation}
which is inspired in the second relation of Eq.  (\ref{moms}).  Except
for  a  dimensionality  factor,  $\sigma^2$  is  thus  the mean square
displacement associated with $p({\bf x})$.

Under these conditions, the maximization of entropy yields
\begin{equation}    \label{Gauss}
p({\bf x}) = (2\pi\sigma^2)^{-d/2}\exp(-x^2/2\sigma^2),
\end{equation}
namely, a Gaussian jump distribution. Since, in view of the constraint
(\ref{const}), the  mean square  displacement associated  with $p({\bf
x})$ is  finite, the  maximum-entropy  formalism applied  as above  to
the jump distribution of a random walk describes normal diffusion.

The question  on whether  anomalous diffusion  can be  derived from  a
variational  formalism  for  the  entropy  arises now quite naturally.
Montroll and  Shlesinger \cite{MS}  have shown  that this  is in  fact
possible, but  requires replacing  the constraint  (\ref{const}) by  a
more complex condition on $p({\bf x})$. In particular, L\'evy flights,
Eq. (\ref{Levy}), are obtained  from the  maximization of  the entropy
(\ref{Sx})   if   the   jump   distribution   satisfies,   along  with
normalization,
\begin{eqnarray}
\int &\ln& \left[ (2\pi)^{-d} \int \exp (-i{\bf k}\cdot
{\bf x}-bk^\gamma)\ d {\bf k} \right] \ p({\bf x})\ d{\bf x}
\nonumber \\
&=&\mbox{constant}.  \label{constMS}
\end{eqnarray}
This is however a quite  unsatisfactory answer to the above  question.
Indeed,  besides  its  complexity,  the  constraint (\ref{constMS}) is
anything but a natural condition  to impose to the jump  distribution.
In Montroll and Shlesinger's words, ``it is difficult to imagine  that
anyone in an {\it a priori} manner would introduce'' such a  condition
for maximizing the entropy with  respect to $p({\bf x})$. This  remark
would at once  exclude L\'evy flights  --and anomalous diffusion  with
them-- from the frame of the maximum-entropy formalism and, therefore,
from a natural connection with equilibrium statistics.

In Ref. \cite{AZ},  a different approach  has been proposed  to tackle
the problem of deriving  anomalous diffusion from the  maximization of
entropy.   Since  replacing  the  constraint  on the jump distribution
implies imposing unconventional, forced conditions on $p({\bf x})$,  a
possible way out is  to replace the form  of the entropy instead.   In
particular, it has been found that the form of the entropy proposed by
Tsallis  \cite{Ts1,Ts15}  produces,   upon  maximization  with   the
constraints  prescribed  by  this  generalized  theory, power-law jump
distributions with the asymptotic behavior given in (\ref{asymp}).  As
described in the following, random-walk models of anomalous  diffusion
find thus a natural statistical-mechanical basis in Tsallis' theory.

\subsection{Generalized formalism}

Inspired in the theory  of multifractals, Tsallis \cite{Ts1}  proposed
to generalize the traditional Boltzmann-Gibbs statistical mechanics by
introducing new forms  for the entropy  and for the  constraints to be
applied in the maximization procedure. For a system whose $i$-th state
is occupied with probability $p_i$, the generalized entropy reads
\begin{equation}    \label{Sq}
S_q[p]=-\frac{1-\sum_i p_i^q}{1-q},
\end{equation}
where $q$ is a real  parameter. For the canonical ensemble,  where the
energy of the $i$-th state  is $\epsilon_i$ and the average  energy is
$E_q$, the generalized constraint to  be imposed to $p_i$, along  with
probability normalization, is
\begin{equation}    \label{constq}
\sum_i \epsilon_i p_i^q = E_q.
\end{equation}
In this generalized formalism, in fact, the average of any  observable
$O$  is  defined  as  $\langle  O  \rangle_q = \sum_i O_i p_i^q$. This
average is  usually refered  to as  the $q$-expectation  value of  $O$
\cite{Ts2}.

The  generalized  statistical-mechanical   formalism  based  on   Eqs.
(\ref{Sq}) and (\ref{constq}) has some remarkable properties. First of
all, it reduces to the traditional Boltzmann-Gibbs formulation in  the
limit $q \to 1$. In fact, Eq. (\ref{Si}) is recovered from  (\ref{Sq})
in  that  limit  except  for  the  factor  $k_B$,  which has here been
conventionally  put  equal   to  unity.    The  canonical   constraint
(\ref{constq}) reduces in turn  to the traditional definition  of mean
energy. The new  formalism preserves the  full Legendre-transformation
structure of thermodynamics for all $q$ \cite{Ts15}, leaving invariant
in form the  main results of  statistical thermodynamics, such  as the
Ehrenfest  theorem,  the  H-theorem,  the  von  Neumann  equation, the
Bogolyubov inequality, and the Onsager reciprocity theorem \cite{Ts2}.
Its seems to be particularly useful in dealing with systems  involving
long-range correlations and  non-extensivity, as the  formalism itself
is non-extensive for $q\neq 1$. The Tsallis exponent $q$ has thus been
interpreted as a measure of non-extensivity. Since its introduction  a
decade  ago  \cite{Tsa}  Tsallis   statistics  has  found   successful
applications to a  large class of  problems of high  interest, ranging
from  gravitational  systems,  to  turbulent  flows,  to  optimization
algorithms.   Many of  these applications  are described  in detail in
other  papers  of  the  present  issue,  and  are  therefore no longer
discussed here.

In order to apply Tsallis statistics to discrete-time random walks  in
the  spirit  outlined  in  the  previous  section, Eqs. (\ref{Sx}) and
(\ref{const})  have  to  be  generalized  according  to (\ref{Sq}) and
(\ref{constq}), respectively. As a  function of the jump  probability,
the generalized entropy can be written as \cite{Tsa,AZ,Ts3}
\begin{equation}    \label{Sqx}
S_q[p] = - \frac{1}{1-q} \left\{ 1- \sigma^{-d}\int
\left[ \sigma^d p({\bf x})\right]^q d{\bf x} \right\},
\end{equation}
whereas the canonical constraint transforms into a condition on the
$q$-expectation value of $x^2$:
\begin{equation}    \label{constqx}
\langle x^2  \rangle_q =\sigma^{-d}\int x^2
\left[\sigma p({\bf x})\right]^q d{\bf x}=\sigma^2.
\end{equation}
Here,  $\sigma$  preserves  its  identification  as  a  typical length
associated  with  the  jump  probability.   However,  for  $q\neq  1$,
$\sigma^2$ does not coincide with the mean square length of the jumps.
For simplicity, the  dimensionality factor in  the right-hand side  of
Eq.  (\ref{const}) has now been absorved by $\sigma$.

It  is  shown  in  the  following  that  the maximization of $S[p]$ as
defined  in  (\ref{Sqx})  with  the  constraints  (\ref{normal})   and
(\ref{constqx}) --which, as in the case of the traditional  formalism,
is  carried  out  by  the  standard  method  of  Lagrange  multipliers
\cite{Ts1,Ts15}-- produces a power-law jump distribution. The exponent
of the  power-law depends  on the  space dimension  and on the Tsallis
exponent $q$.   This jump  distribution is  not a  L\'evy distribution
like (\ref{Levy}), but has the  same type of asymptotic behavior,  Eq.
(\ref{asymp}).   For  suitable  values  of  $d$  and $q$, this form of
$p({\bf  x})$  will  therefore  define  a  random  walk with anomalous
properties.

For the sake of clarity, the results in one dimension are shown  first
\cite{Tsa,Ts3}.  The jump distribution resulting from the maximization
procedure is, in this case,
\begin{equation}    \label{p1}
p(x)  = {\cal Z}_q^{-1}\left[ 1-\beta(1-q)x^2 \right]^{1/(1-q)}
\end{equation}
where the partition function ${\cal Z}_q$ is given by
\begin{equation}    \label{Z1}
{\cal Z}_q= \int_{-\infty}^{+\infty}
\left[ 1-\beta(1-q)x^2 \right]^{1/(1-q)} dx.
\end{equation}
The  positive  constant  $\beta$  is  one of the Lagrange multipliers,
which can be expressed as a function of $\sigma$ using the  constraint
(\ref{constqx}), as shown  below.  In  the generalized formulation  of
statistical mechanics $\beta$ is related to the temperature $T$ in the
standard form, $\beta \propto 1/T$.

\begin{figure}
\begin{center}
\psfig{figure=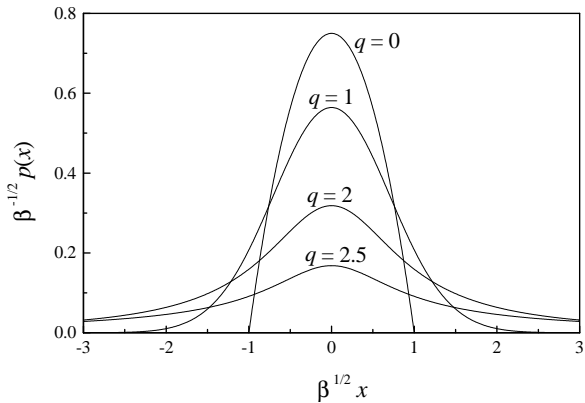,width=\columnwidth}
\end{center}
\caption{Jump   probability   distribution   derived   from    Tsallis
statistics, for some values of the Tsallis exponent $q$. For $q=1$ the
standard Gaussian  is obtained.  Note the  cut-off for  $q<1$, and the
long-tailed power-law distributions for $q>1$.}
\label{f3}
\end{figure}

It turns out from Eq.  (\ref{p1}) that the normalization of $p(x)$ can
be satisfied for $q<3$ only.  The Tsallis exponent for random walks is
therefore  restricted  to  the  interval  $(-\infty,3)$.    Explicitly
calculating the partition function yields
\begin{equation}    \label{p11}
p(x)=\sqrt{\frac{\beta(q-1)}{\pi}}
\frac{\Gamma\left(\frac{1}{q-1}\right)}
{\Gamma\left(\frac{1}{q-1}-\frac{1}{2}\right)}
\left[ 1+\beta(q-1)x^2 \right]^{-1/(q-1)}
\end{equation}
for $1<q<3$, and
\begin{equation}    \label{p12}
p(x)=\left\{
\begin{array}{ll}
\sqrt{\frac{\beta(1-q)}{\pi}}
\frac{\Gamma\left(\frac{1}{1-q}+\frac{3}{2}\right)}
{\Gamma\left(\frac{1}{1-q}+1\right)}
&\left[ 1-\beta(1-q)x^2 \right]^{1/(1-q)} \\
&\mbox{if $x^2<{1}/{\beta(1-q)}$}, \\ \\
0  &\mbox{otherwise}
\end{array}
\right.
\end{equation}
for $q<1$.   For $q\to  1$, of  course, the  Gaussian (\ref{Gauss}) is
reobtained.   Note that  the Tsallis  exponent $q$  is related  to the
exponent $\gamma$ in Eq. (\ref{asymp}) according to \cite{AZ}
\begin{equation}    \label{gq}
\gamma=\frac{3-q}{q-1}\ \ \ \ \ \mbox{or} \ \ \ \ \
q=\frac{3+\gamma}{1+\gamma}  .
\end{equation}
Figure \ref{f3}  shows the  profile of  $p(x)$ for  several values  of
$q$.

Regarding anomalous diffusion, thus, it  is clear that the case  $q<1$
is  irrelevant.   In  fact,  for  such  values of the Tsallis exponent
$p(x)$ exhibits a cut-off at $|x|= 1/\sqrt{\beta(1-q)}$, and  vanishes
for larger $|x|$, as shown by Eq.  (\ref{p12}).  This implies at  once
that the mean square displacement associated with $p(x)$ is finite and
the  resulting  random  walk  corresponds  to  normal  diffusion.  The
attention  is  consequently  focused  in  the  following  on  the case
$1<q<3$, Eq. (\ref{p11}).  In this case, the mean square  displacement
is
\begin{equation}    \label{msd1}
\langle x^2 \rangle =\left\{
\begin{array}{ll}
[\beta(5-3q)]^{-1} & \mbox{if $q<5/3$,}\\ \\
\infty & \mbox{if $q\geq 5/3$}.
\end{array}
\right.
\end{equation}
Therefore, for $1<q<5/3 \approx 1.67$ the mean square displacement  is
still finite, and the random walk corresponds to normal diffusion.  On
the other hand, anomalous superdiffusion  is obtained for $5/3 \leq  q
<3$.

It is interesting to calculate now the $q$-expectation value of  $x^2$
which, in the  frame of Tsallis  statistics, replaces --as  an average
quantity-- the mean square  displacement of the standard  formulation.
According to the constraint  (\ref{constqx}) imposed to $p(x)$  in the
maximization of entropy, this $q$-expectation value should be  finite.
In fact,
\begin{equation}    \label{qexp}
\langle x^2 \rangle_q =
\frac{1}{2\beta}\left[
\sqrt{\frac{(q-1)}{2\pi}}
\frac{\Gamma\left(\frac{1}{q-1}\right)}
{\Gamma\left(\frac{1}{q-1}-\frac{1}{2}\right)}
\right]^{2(q-1)/(3-q)},
\end{equation}
for  $1<q<3$.   The  fact  that,  in  contrast  with  the  mean square
displacement, $\langle  x^2 \rangle_q$  is finite,  seems to  indicate
that the constraint (\ref{constqx}) is  a natural one in the  frame of
anomalous-diffusion  random  walks  \cite{AZ}.  Note moreover that Eq.
(\ref{qexp}) along with  (\ref{constqx}) gives the  connection between
the  Lagrange  multiplier  $\beta$   and  the  characteristic   length
$\sigma$,
\begin{equation}    \label{bs}
\beta \propto \sigma^{-2}.
\end{equation}

\begin{figure}
\begin{center}
\psfig{figure=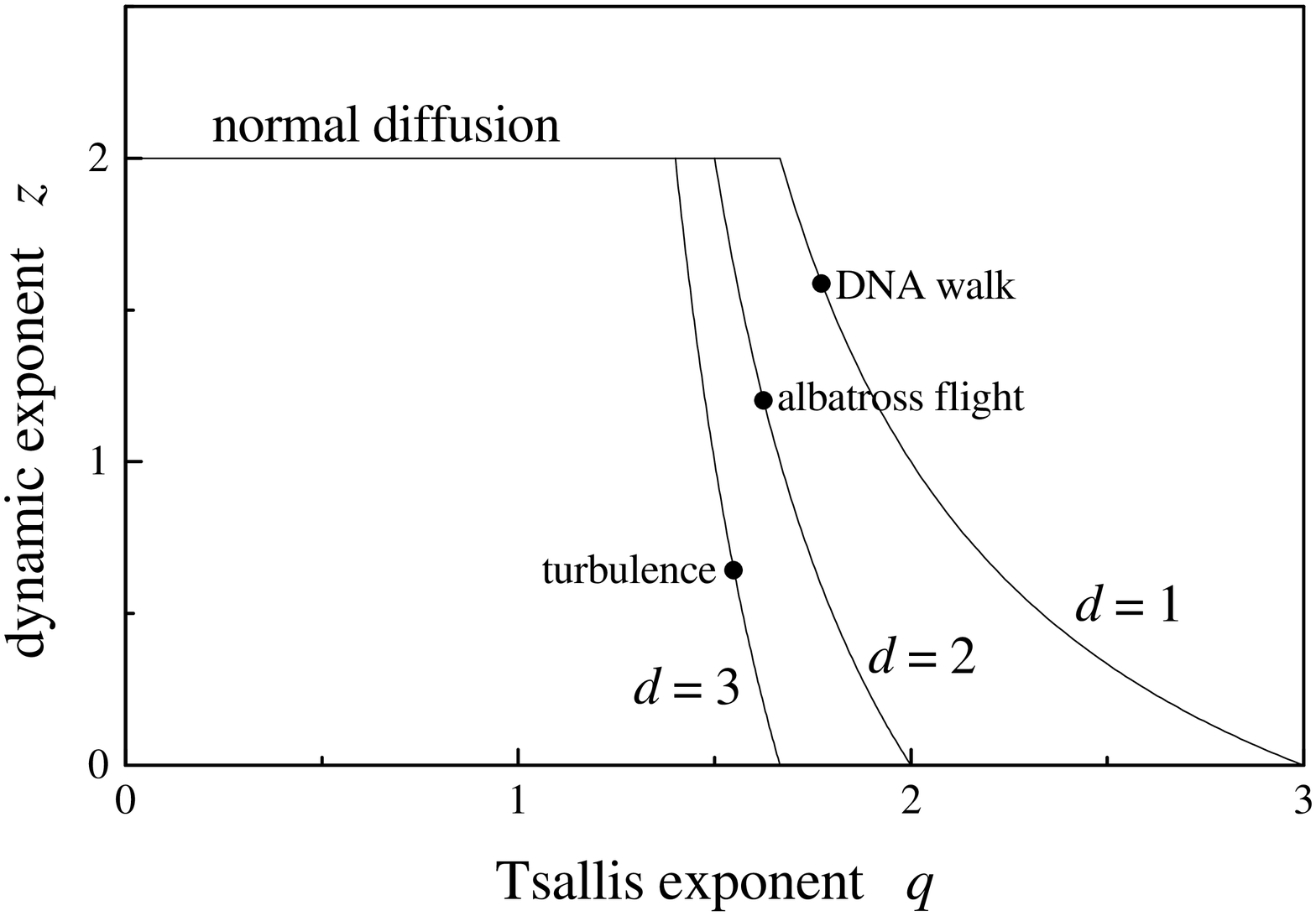,width=\columnwidth}
\end{center}
\caption{The  diffusion  dynamic  exponent  $z$  as  a function of the
Tsallis exponent $q$, for different spatial dimensions. The dots
stand for some of the instances of anomalous diffusion discussed in
Section \ref{sect2.1}.}
\label{f4}
\end{figure}

Equations  (\ref{p11})  and  (\ref{p12})  make  clear  that, as stated
above, the maximization of entropy within Tsallis' formalism does  not
lead to L\'evy distributions for the jump probability. Rather, a plain
power-law  function  of  the  jump  length  $x$  is  obtained.  L\'evy
distributions  are  however  reobtained  when considering the temporal
evolution  of  the  random  walk  generated  by  $p(x)$.  In fact, the
displacement $r$ of the  walker after $t$ time  steps is given by  the
sum of  the successive  jumps. By  virtue of  the generalized  central
limit theorem  \cite{Lev,CLT} discussed  in Section  \ref{sect2.2} the
probability distribution  $P(r,t)$ is  thus given,  for large  $t$ and
$\gamma<2$ (i.e. $q\geq  5/3$), by a  stable L\'evy distribution  with
L\'evy exponent $\gamma$. For $\gamma>2$, on the other hand, the usual
form of  the central  limit theorem  holds and  the total displacement
distribution is a  Gaussian.  The  dynamic exponent $z$  of the random
walk  --which  coincides  with  $\gamma$  for  $\gamma<2$ (see Section
\ref{sect2.2})-- is then
\begin{equation}    \label{zq}
z =\left\{
\begin{array}{ll}
2 & \mbox{if $q<5/3$,}\\ \\
(3-q)/(q-1) & \mbox{if $q\geq 5/3$}.
\end{array}
\right.
\end{equation}
This connection between the dynamic exponent and the Tsallis  exponent
$q$ --which  is illustrated  by the  curve $d=1$  in Figure \ref{f4}--
constitutes indeed  the main  result of  the description  of anomalous
diffusion in the frame of the  Tsallis' formulation.  It shows that  a
close  relation  exists  between  the  properties  of  a L\'evy-flight
process and the non-extensiveness of the involved statistics.  As  far
as the  underlying statistical  frame differs  from  Boltzmann-Gibbs',
the  maximum-entropy  formalism  produces  a  random walk which models
superdiffusion as a L\'evy flight.

As  in  the  case  of  the  jump  distribution,  the total mean square
displacement  $\langle  r^2  \rangle$  associated  with L\'evy flights
diverges. On  the other  hand, the  $q$-expectation value  of $r^2$ is
well defined for  all relevant $q$  ($1<q<3$), cf.   Eq. (\ref{qexp}).
This can be calculated taking  into account the scaling properties  of
$P(r,t)$, Eq. (\ref{scaling}), and reads
\begin{equation}    \label{rq}
\langle r^2 \rangle_q = \left\{
\begin{array}{ll}
D(q)\beta^{-1}t^{(3-q)/2} & \mbox{if $q<5/3$,}\\ \\
D(q)\beta^{-1}t^{q-1} & \mbox{if $q\geq 5/3$}.
\end{array}
\right.
\end{equation}
The proportionality  factor $D(q)$  depends on  $q$ only. Interpreting
now the Lagrange multiplier $\beta$ as the inverse of the  temperature
--as prescribed in the  frame of Tsallis thermodynamics  \cite{Ts15}--
the above  equation can  be seen  as a  generalization of the Einstein
relation  (\ref{Einstein})  \cite{ZA}.   Equations  (\ref{width})  and
(\ref{Einstein})  imply  in  fact  that  $\langle  r^2\rangle  \propto
\beta^{-1}t$  for  normal  diffusion,  and  Eq.   (\ref{rq})  is   the
extension of this result to Tsallis statistics.  Again, the fact  that
$\langle  r^2\rangle_q$  is  finite  for  L\'evy flights suggests that
Tsallis'  formalism  provides  a  natural  frame  for  the statistical
description of such kind of anomalous diffusion.

\begin{figure}
\begin{center}
\psfig{figure=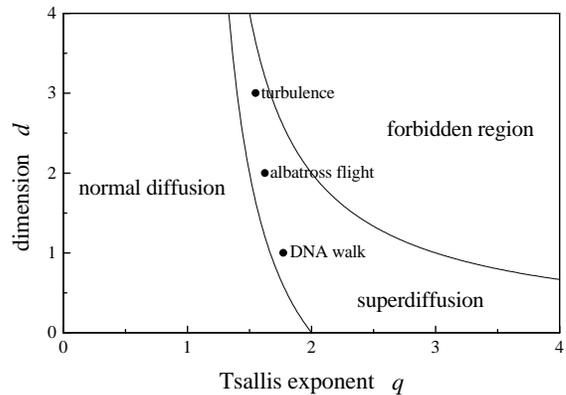,width=\columnwidth}
\end{center}
\caption{Phase diagram in the  ($q,d$)-plane, displaying the zones  of
normal and anomalous  diffusion, and the forbidden region where  the
jump probability distribution cannot be normalized. The dots
stand for some of the instances of anomalous diffusion discussed in
Section \ref{sect2.1}.}
\label{f5}
\end{figure}

Though the algebra  is more involved  than above, anomalous  diffusion
in more  than one  dimension can  be straightforwardly  treated in the
frame  of   Tsallis  statistics,   and  the   main  conclusions    are
qualitatively the same as for the one-dimensional case. Maximazing the
entropy given  in (\ref{Sqx})  in the  $d$-dimensional space  produces
formally  the  same  jump  distribution  as  in  (\ref{p1}), where the
partition function has however  to be calculated as  a $d$-dimensional
integral. The jump probability can be normalized if
\begin{equation}    \label{q1}
q<\frac{2+d}{d},
\end{equation}
and the associated mean square displacement is finite if
\begin{equation}    \label{q2}
q<\frac{4+d}{2+d}.
\end{equation}
Below  this  value,  thus,  the  random  walk  generated  by  the jump
probability  models  normal  diffusion,  whereas  for $(4+d)/(2+d) <q<
(2+d)/2$ the walker performs superdiffusion. In Figure \ref{f5}  these
different regimes are  identified in a  phase diagram. The  connection
between the dynamic exponent and the Tsallis exponent reads now
\begin{equation}    \label{zqd}
z =\left\{
\begin{array}{ll}
2 & \mbox{if $q<(4+d)/(2+d)$,}\\ \\
2/(q-1)-d & \mbox{if $(4+d)/(2+d)<q<(2+d)/d$}.
\end{array}
\right.
\end{equation}
This connection is represented graphically in Figure \ref{f4}.

\section{Nonlinear diffusion and Tsallis statistics}

The diffusion equation,
\begin{equation} \label{diffusion1}
\frac{\partial P}{\partial t}=D \nabla^2_{\bf r} P,
\end{equation}
which governs the  evolution of the  probability $P({\bf r},t)\  d{\bf
r}$ of  finding a  Brownian particle  in a  neighborhood $d{\bf r}$ of
point $\bf r$  at time $t$,  can be generalized  to take into  account
additional mechanisms acting on  both the microscopic dynamics  of the
particle and the mesoscopic dynamics of an ensemble of such particles.
A rather  straightforward generalization  introduces for  instance the
effect of  an external  force field,  ${\bf F}({\bf  r},t)$, acting on
each particle  \cite{VK,Wio}. This  force field  enters the  diffusion
equation as a drift term, namely,
\begin{equation} \label{FP}
\frac{\partial P}{\partial t}=-\nabla_{\bf r}
\cdot ({\bf F}P)+D \nabla^2_{\bf r} P.
\end{equation}
This equation, which  combines the effect  of probability drift  --due
to the force-- and of  probability spreading --due to diffusion--  can
be seen to govern a huge class of random processes, in a generic space
of states  $\bf r$  \cite{VK,Wio}. It  is generally  refered to as the
Fokker-Planck equation.

Further generalizations, mainly justified on a phenomenological basis,
have  lead  to  propose  a  nonlinear  version  of  the  Fokker-Planck
equation \cite{LB}, namely,
\begin{equation} \label{NLFP}
\frac{\partial P^\mu}{\partial t}=-\nabla_{\bf r}
\cdot ({\bf F}P^\mu)+D \nabla^2_{\bf r} P^\nu,
\end{equation}
where $\mu>0$ and $\nu$ are suitable real constants.  Without  loosing
generality,  one  can  fix  $\mu=1$,  by  changing  $P^\mu  \to P$ and
$\nu/\mu\to  \nu$.    In   such  case,   Eq.    (\ref{NLFP})  can   be
phenomenologically interpreted as  a Fokker-Planck equation  where, if
$\nu\neq  1$,  the  diffusion  coefficent  depends  on the probability
$P({\bf  r},t)$.    This   density-dependent  diffusivity   represents
nonlinear effects arising, for instance, from interaction between  the
diffusing particles. Such kind of nonlinearities have been observed in
several real processes,  such as transport  in porous media  ($\nu \ge
2$) \cite{porous}, surface  growth ($\nu=3$) \cite{surf},  liquid film
spreading under gravity ($\nu=4$) \cite{films}, and Marshak  radiative
heat transfer ($\nu=7$) \cite{Mar}, among others \cite{others}.

The Fokker-Planck equation (\ref{FP}) is linear. This implies that for
many forms of the force  field ${\bf F}({\bf r},t)$ the  general exact
solution  can  be  found  analytically.  Moreover,  even if analytical
solutions are  not available  or difficult  to obtain,  very efficient
computational  methods   can  be   implemented  to   obtain  numerical
solutions.  In  contrast,  exact  solutions  to the nonlinear equation
(\ref{NLFP})  are  particularly   scarce  \cite{scarce,scarce2},   and
numerical methods are typically subject to instabilities when  dealing
with nonlinear problems. It is  therefore of great interest to  verify
that, as shown in the following, Tsallis' formalism provides a  method
to obtain special solutions to Eq. (\ref{NLFP}) for general $\mu$  and
$\nu$, at least, in its  one dimensional version and for  some special
forms of the  drift force. This  problem has recently  been treated by
Tsallis himself \cite{TsB}  and, in an  alternate form, by  Compte and
Jou \cite{Jou}.

Consider then the one dimensional version of Eq. (\ref{NLFP}) for  the
probability density $P(x,t)$,
\begin{equation} \label{FP1}
\frac{\partial P^\mu}{\partial t}=-\frac{\partial}{\partial x} [
F(x)P^\mu ]+D \frac{\partial^2 P^\nu}{\partial x^2},
\end{equation}
with  $x\equiv  r$,  and  with $F(x)=k_1-xk_2$. This time-independent,
linear form of the drift force corresponds, in general, to a quadratic
potential  --i.e.  an  Ornstein-Uhlenbeck  random  process \cite{VK}--
whereas it reduces to a linear potential --namely, a constant  force--
for $k_2=0$.  For $\mu=\nu=1$, this equation  can be straightforwardly
solved,  for  instance,  by  Fourier-Laplace  transforming.  A special
solution is
\begin{equation}    \label{P}
P(x,t)=\frac{1}{{\cal Z}(t)}\exp\left\{ \beta(t)[x-\bar x(t)]^2
\right\},
\end{equation}
where
\begin{equation}    \label{P1}
\frac{\beta(t)}{\beta(0)}= \left[ \frac{{\cal Z}(0)}{{\cal
Z}(t)} \right]^2 =\left[(1-\Delta) \exp(-2k_2 t)+\Delta
\right]^{-1}
\end{equation}
with $\Delta=2D\beta(0)/k_2$, and
\begin{equation}    \label{P2}
\bar x(t)=\kappa +[\bar x(0)-\kappa]\exp(-k_2t)
\end{equation}
with  $\kappa=k_1/k_2$.  This  particular  solution  has the important
property that, for $t\to 0$ and $\beta(0)\to \infty$, it reduces to  a
delta-like distribution,  $P(x,0) =  \delta[x-\bar x(0)]$.   Since Eq.
(\ref{FP1}) is linear for $\mu=\nu=1$, and delta distributions can  be
used  as  a  base  for  the  space  of  initial conditions $P(x,0)$, a
suitable  linear  combination  of  functions  of  the  form  (\ref{P})
provides the solution to the linear equation for any initial condition
in that space. In this sense, (\ref{P}) gives the general solution  to
the  linear  Fokker-Planck  equation  with  the above prescribed drift
force.

Focus  now  the  attention  on  the  functional form of the particular
solution to the linear problem  given in Eq. (\ref{P}). As  a function
of  $x$,  $P(x,t)$  is  a  Gaussian,  essentially  of the same type as
(\ref{Gauss}).  The  differences  are,   firstly,  that  the   spatial
coordinate $x$ is shifted by an  amount $\bar x(t)$.  The Gaussian  is
therefore centered around a position which depends on time.  Secondly,
the width of  the Gaussian, which  is proportional to  $\beta^{-1/2}$,
depends  also  on  time.   Since  the  solution  (\ref{P})   preserves
normalization,   the   normalization   factor   ${\cal   Z}^{-1}$   is
time-dependent.

The Gaussian profile of $P(x,t)$  in Eq. (\ref{P}) suggests that  this
solution  can   be  formally   derived  from   a  suitably    extended
maximum-entropy formalism, in its standard Boltzmann-Gibbs version. In
fact, it can  be shown \cite{TsB}  that such form  of $P(x,t)$ derives
from the maximization of
\begin{equation}    \label{SP}
S[P] =\int_{-\infty}^{+\infty}P(x,t) \ln P(x,t)\ dx,
\end{equation}
with the extended constrains
\begin{equation}    \label{C1}
\int_{-\infty}^{+\infty} P(x,t)\ dx =1,
\end{equation}
\begin{equation}    \label{C2}
\int_{-\infty}^{+\infty} [x-\bar x(t)]P(x,t)\ dx =0,
\end{equation}
and
\begin{equation}    \label{C3}
\int_{-\infty}^{+\infty} [x-\bar x(t)]^2 P(x,t)\ dx
=\frac{1}{2\beta(t)},
\end{equation}
for arbitrary $\bar x(t)$ and  $\beta(t)$. The special forms of  these
functions that  make the  probability distribution  satisfy the linear
Fokker-Planck equation can be  obtained by simply replacing   $P(x,t)$
in the equation.

\begin{figure}
\begin{center}
\psfig{figure=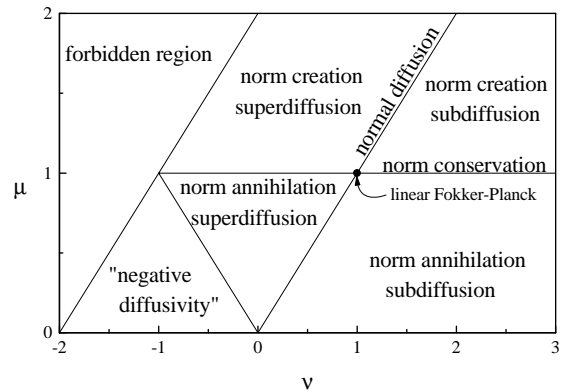,width=\columnwidth}
\end{center}
\caption{Phase  diagram  in  the  ($\nu,\mu$)-plane,  displaying   the
different regimes  of anomalous  diffusion and  norm evolution  in the
solutions to the nonlinear Fokker-Planck equation. In the forbidden
region the probability distribution cannot be normalized.}
\label{f8}
\end{figure}

It should be by now clear that one is immediately interested at  which
solutions  are  obtained  if,  instead  of  the  standard maximization
principle,  the  Tsallis'   formalism  is  used.   Namely,  take   the
generalized entropy
\begin{equation}    \label{SPP}
S_q[P] =-\frac{1}{1-q}
\left\{ 1- \int_{-\infty}^{+\infty} P(x,t)^q dx\right\} ,
\end{equation}
and  maximize  it  with  respect  to $P(x,t)$ imposing the generalized
constrains
\begin{equation}    \label{Cq2}
\int_{-\infty}^{+\infty} [x-\bar  x(t)]P(x,t)^q dx =0,
\end{equation}
and \begin{equation}    \label{Cq3}
\int_{-\infty}^{+\infty} [x-\bar x(t)]^2 P(x,t)^q dx
=\frac{1}{2\beta(t)}.
\end{equation}
This produces \cite{TsB}
\begin{equation}    \label{Pq}
P(x,t)= \frac{1}{{\cal Z}_q(t)} \left\{  1-\beta(t) (1-q)[x-\bar x
(t)]^2\right\}^{1/(1-q)},
\end{equation}
to be compared with Eq.   (\ref{p1}). Remarkably enough, this form  of
$P(x,t)$ turns out to be a solution  to the  one-dimensional nonlinear
Fokker-Planck equation for the linear drift force if
\begin{equation}    \label{qmunu}
q=1+\mu-\nu,
\end{equation}
and
\begin{equation}    \label{betaZ}
\frac{\beta(t)}{\beta(0)}= \left[ \frac{{\cal Z}_q(0)}{{\cal
Z}_q(t)}\right]^{2\mu}.
\end{equation}
As for L\'evy-flight anomalous  diffusion, $P(x,t)$ can be  normalized
only if $q<3$.  This defines a forbidden region for  $\mu>2+\nu$ (Fig.
\ref{f8}).

\begin{figure}
\begin{center}
\psfig{figure=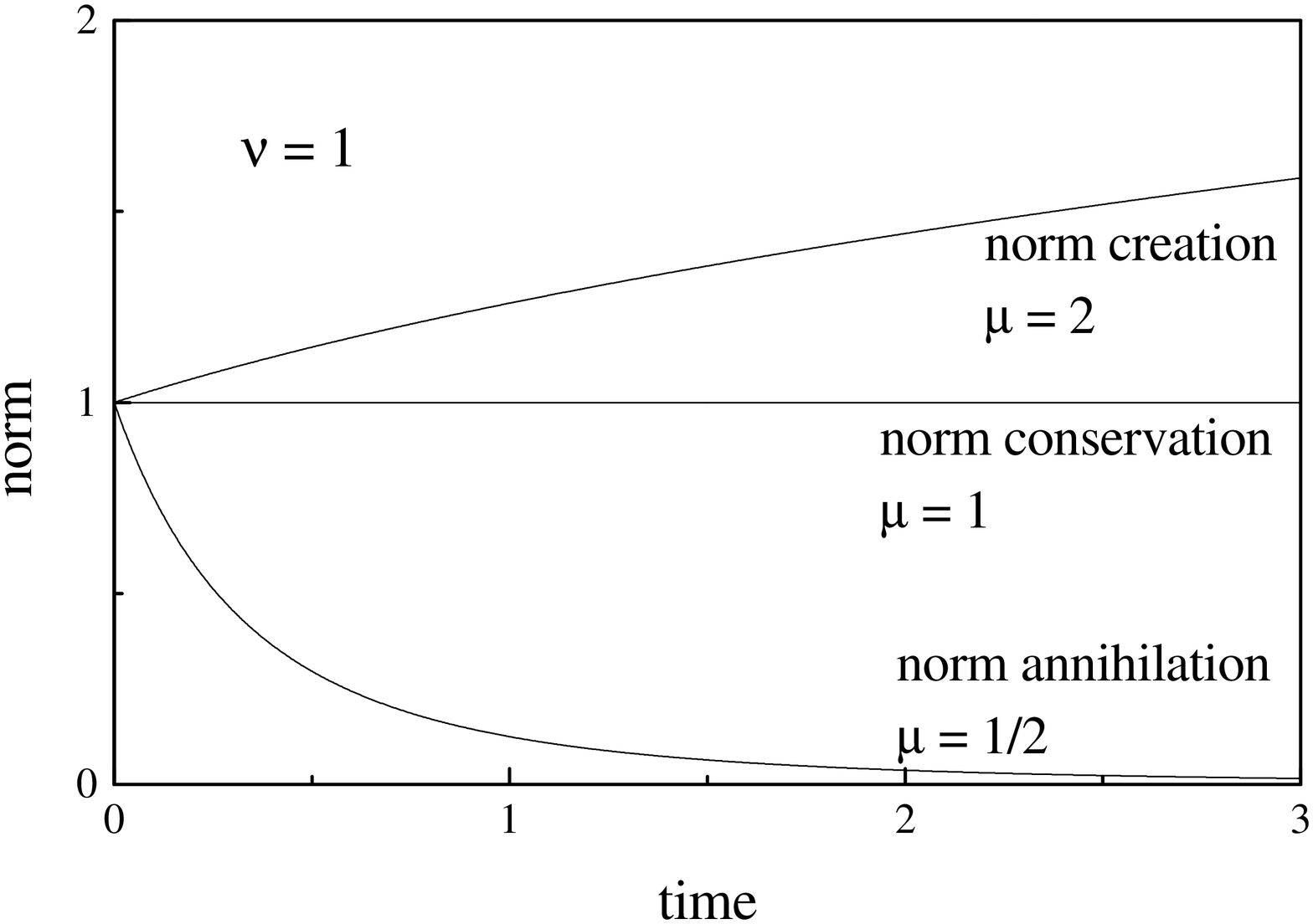,width=\columnwidth}
\end{center}
\caption{Evolution of the norm in the solutions to the nonlinear
Fokker-Planck equation for $\nu=1$ and some values of $\mu$.
These curves correspond to $\Delta_q>1$.}
\label{f6}
\end{figure}

The function ${\cal Z}_q (t)$ is explicitely given by
\begin{equation}    \label{Z(t)}
{\cal Z}_q (t)=
{\cal Z}_q (0)\left[(1-\Delta_q) \exp(-t/\tau)+\Delta_q
\right]^{1/(\mu+\nu)},
\end{equation}
with
\begin{equation}    \label{Deltaq}
\Delta_q=\frac{2\nu D\beta(0) {\cal Z}_q(0)^{\mu-\nu}}{k_2},
\end{equation}
and $\tau=\mu/k_2(\mu+\nu)$. The function $\bar x (t)$ is the same  as
for the  linear case,  Eq.   (\ref{P2}).   Note that the normalization
constraint, Eq. (\ref{C1}), has  not been imposed in  the maximization
of $S_q[P]$. In fact, the preservation of the norm of $P(x,t)$ is  now
not compatible with  the other two  constraints. Rather, it  turns out
that the  integral of  the probability  density over  the whole  space
varies with time according to
\begin{equation}    \label{normt}
\int_{-\infty}^{+\infty} P(x,t)\ dx =
\left[ \frac{{\cal Z}_q(t)}{{\cal Z}_q(0)}\right]^{\mu-1}
\int_{-\infty}^{+\infty} P(x,0)\ dx.
\end{equation}
This implies that the norm is conserved for all times only if $\mu=1$,
or if  $\Delta_q=1$ --when  ${\cal Z}_q$  does not  depend on time. If
$\Delta_q>1$  the  norm  monotonically   increases  for  $\mu>1$   and
decreases  for  $\mu<1$.   If  $\Delta_q  <1$,  on the other hand, the
opposite behavior is observed. Moreover, for $\nu<0$ the norm diverges
or  vanishes  at  a  finite  time.  Figure  \ref{f6} illustrates these
different regimes for $\nu=1$ and some values of $\mu$.

\begin{figure}
\begin{center}
\psfig{figure=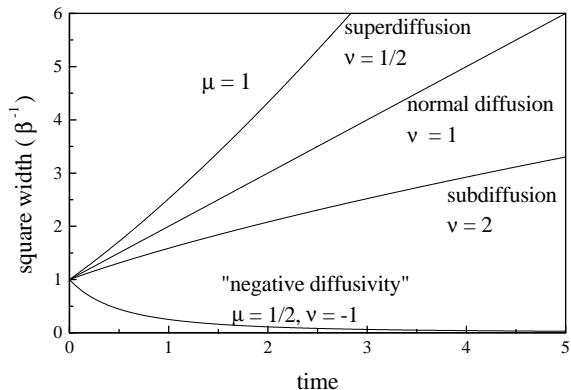,width=\columnwidth}
\end{center}
\caption{Evolution of the width in the solutions to the nonlinear
Fokker-Planck equation for $\mu=1$ and some values of $\nu$.
For the case of ``negative diffusivity,'' $\mu=1/2$.}
\label{f7}
\end{figure}

The case of constant force,  $k_2=0$, can be analyzed as  the suitable
limit of  the above  solution for  $k_2 \to  0$. In particular, taking
$\exp(-t/\tau) \approx 1-t/\tau$ in Eq. (\ref{Z(t)}) it is found that
\begin{equation}    \label{Z11}
{\cal Z}_q (t)=
{\cal Z}_q (0)\left[1+2\frac{\nu(\nu+\mu)}{\mu}D \beta(0) {\cal Z}_q
(0)^{\mu-\nu} t \right]^{1/(\mu+\nu)}.
\end{equation}
In this limit, the width  of the distribution --which is  proportional
to $\beta^{-1/2}$--  exhibits a  well-defined power-law  dependence on
time.  In  fact,  according  to  Eq.  (\ref{betaZ}),  $\beta(t)^{-1/2}
\propto t^{\mu/(\nu+\mu)}$.  This makes  possible to  assign a dynamic
exponent to this kind of diffusion, given by
\begin{equation}    \label{zFP}
z=1+\frac{\nu}{\mu}.
\end{equation}
Whereas, as expected, the case $\mu=\nu=1$ corresponds thus to  normal
diffusion, $\nu/\mu >1$ corresponds to subdiffusion and $\nu /  \mu<1$
corresponds to  superdiffusion. Note  that if  $\nu/\mu<-1$, $z<0$ and
the distribution width decreases with time. This seemingly  unphysical
situation  \cite{TsB}  can  however  be  associated  with  a  kind  of
``negative  diffusivity'',  which  has  been  observed  in  some  real
nonequilibrium self-organizing systems \cite{scarce}.  Figure \ref{f7}
shows the evolution of the distribution width for $\mu=1$ and  various
values of $\nu$.  The phase diagram of Figure \ref{f8} summarizes  the
various   regimes    obtained   in    different   regions    of    the
($\nu,\mu$)-plane.

Equation (\ref{qmunu}) makes evident that the non-extensivity inherent
to Tsallis statistics is related,  in the frame of its  application to
the resolution of the nonlinear Fokker-Planck equation (\ref{FP1}), to
the nonlinearity of the equation itself. This nonlinearity translates,
at  the  level  of  the  solutions,  into  anomalous properties of the
involved  transport  processes.   Thus,  a  clear  connection  between
anomalous  diffusion  and  the   non-extensivity  of  the   underlying
statistics  arises  again.   It  is  important  to  point out that, in
contrast with Eq. (\ref{P}), the solutions (\ref{Pq}) to the nonlinear
Fokker-Planck  equation  derived  from  Tsallis'  formalism  cannot be
combined to give a general solution. In fact, due to the  nonlinearity
of Eq. (\ref{FP1}), no  superposition principle holds, and  (\ref{Pq})
are  particular  solutions  for   special  initial  conditions   only.
Nevertheless, the  straightforward way  in which  these solutions have
appeared  as  an  extension  of  the  linear  case  along the lines of
Tsallis'  generalization,  reinforces  strongly  the  close   relation
between Tsallis' formalism an anomalous diffusion.

\section{Conclusion}

Though normal diffusion is ubiquitous  in Nature, a large --and  still
growing--  class  of  real  systems  is  driven by a different kind of
transport processes, namely, by anomalous  diffusion.  In view of  the
current  importance  of  many  of  these  systems  --which  range from
turbulent flows, to disordered  media and chaotic dynamics,  to flight
patterns  in  birds--  it  is  of  high  interest  having  at  hand  a
formulation    able    to    place    anomalous    diffusion    in   a
statistical-thermodynamical frame, generalizing thus Einstein's theory
for  normal  diffusion.   However,  within  Boltzmann-Gibbs statistics
``the wonderfull world of clusters and intermittencies and bursts that
is associated with L\'evy distributions would be hidden from us if  we
depended  on  a  maximum   entropy  formalism  that  employed   simple
traditional auxiliary conditions'' \cite{MS}.  It has been here  shown
that, instead, Tsallis generalized statistics is a strong candidate to
succesfully yield such a formulation.

Tsallis  statistics  provides  a  natural  frame  for the mathematical
foundations of anomalous diffusion in two  forms. In the  first place,
jump   distributions   of    random-walk   models   for    L\'evy-like
superdiffusion can be straightforwardly derived from a maximum-entropy
principle within the generalized  theory. In fact, such  distributions
exhibit power-law long  tails, which are  an essential feature  in the
results  of  the  theory.  At  once,  Tsallis  statistics furnishes an
elegant  explanation  for  the  appearence  of L\'evy distributions in
other natural phenomena, as the result of the superposition of  random
variables with long-tailed  distributions.  In  the second place,  the
functional form of the distributions resulting from Tsallis' formalism
successfully  suggests  the  solution  to  the nonlinear Fokker-Planck
equation, which describes both subdiffusion and superdiffusion.

In  order  to  widen  the  applications  of  Tsallis'  theory  to  the
statistical  foundations  of  anomalous  diffusion,  further  research
should focus on  some problems that  still wait to  be treated in  the
frame  of  the  generalized  statistics.   For  instance,  it would be
important to extend the derivation of random-walk models of  anomalous
diffusion  to  the  case  of  subdiffusion.  As  explained  in Section
\ref{sect2},   this   requires   introducing   suitable   waiting-time
densities. Since  power-law functions  can fulfill  this role, Tsallis
statistics is again a natural stating point to derive such  densities.
Another extension would  regard diffusion processes  on fractals.   In
fact,  L\'evy-like   diffusion  anomalies   are  the   consequence  of
mechanisms  driving  the  dynamics  of  the  diffusing  particles.  An
alternative formulation, which is relevant to many applications, takes
into  account  that  such  anomalies  originate  rather in the complex
geometry  of  the  medium  where  particles  diffuse.   The connection
between fractal  geometry and  Tsallis statistics  has been identified
early,  and  it  can  thus  be  expected  that  diffusion  on  fractal
substrates finds a  satisfactory statistical-mechanical frame  in such
theory.  Finally, it would  be interesting to descend a  level further
in  the  dynamical  bases  of  anomalous  diffusion,  and try to apply
Tsallis'  formalism  to  the  formulation  of deterministic mechanical
approaches to this kind of transport.

As a final remark it is worth mentioning that, very recently, Tsallis'
formalism  has  been  improved  by  redefining  the  normalization  of
$q$-expectation values \cite{Role}. This has solved, in a single step,
two  main  drawbacks  of  the  theory.   In  fact,  in  its   original
formulation,  Tsallis  statistical  mechanics  is  not invariant under
energy shifts, and the $q$-expectation value of a constant depends  on
the state of  the system under  study. Although the  correction to the
theory does not involve important changes in the qualitative  results,
it  represents  a  major  improvement  from  a formal viewpoint. Here,
Tsallis  statistics  has  been  applied  to anomalous diffusion in its
original form.   A relevant  step forward  would be  to reanalyze this
process in the frame of the corrected theory.

\section*{Acknowledgements}

Fruitful  discussions  with  P.A.  Alemany  and  C. Tsallis are warmly
acknowledged.   G.  Drazer  contributed  some  valuable remarks on the
manuscript.  The   author  is   grateful  to   Fundaci\'on  Antorchas,
Argentina, for financial support.


\begin{references}
%
\bibitem{Einst} A. Einstein, Ann. Phys. {\bf 17}, 549 (1905).

\bibitem{MW}  E.W.  Montroll  and  B.J.  West,  in  {\it   Fluctuation
Phenomena},  E.W.   Montroll  and   J.L.  Lebowitz,   eds.  (Elsevier,
Amsterdam, 1979) p. 61.

\bibitem{VK} N.G.  van Kampen,  {\it Stochastic  Processes in  Physics
and Chemistry} (North-Holland, Amsterdam, 1992).

\bibitem{Kubo} R. Kubo, {\it Statistical Mechanics} (North-Holland,
Amsterdam, 1988).

\bibitem{anom} J.-P. Bouchaud and A. Georges, Phys.  Rep. {\bf 195},
127 (1990).

\bibitem{ber} J. Bernasconi {\it et al.}, Phys. Rev. Lett.
{\bf 42}, 819 (1979).

\bibitem{mac} J. Machta, J. Phys. A {\bf 18}, L531 (1985).

\bibitem{Richard}  L.F.  Richardson,  Proc.  Roy.  Soc. London, Ser. A
{\bf 110}, 709 (1926).

\bibitem{Mandel} B.B. Mandelbrot, J. Fluid Mech. {\bf 62}, 331 (1974).

\bibitem{df}  R.A.  Antonia,  N.  Phan-Thien,  and B.R. Satyoparakash,
Phys. Fluids {\bf 24}, 554 (1981).

\bibitem{turbul} M.F.  Shlesinger, B.J.  West, and  J. Klafter,  Phys.
Rev. Lett {\bf 58}, 1100 (1987).

\bibitem{Qflow}  G.M.  Zaslavsky,  R.Z.  Sagdeev,  and A.A. Chernikov,
Sov. Phys.  JETP {\bf  67}, 270  (1988); A.A.  Chernikov {\it et al.},
Phys. Lett. A {\bf 144}, 127 (1990).

\bibitem{Nat} M.F. Shlesinger, G.M. Zaslavski, and J. Klafter,  Nature
{\bf 363}, 31 (1993).

\bibitem{TC}  E.R.  Weeks  {\it  et  al.},  in {\it L\'evy Flights and
Related Topics in  Physics}, M.F. Shlesinger,  G.M. Zaslavsky, and  U.
Frisch, eds. (Springer, Berlin, 1995) p. 51.

\bibitem{DNA} C.-K. Peng {\it et al.}, Nature {\bf 356}, 168 (1992).

\bibitem{DNA1} H.E. Stanley {\it et  al.}, in {\it L\'evy Flights  and
Related Topics in  Physics}, M.F. Shlesinger,  G.M. Zaslavsky, and  U.
Frisch, eds. (Springer, Berlin, 1995) p. 331.

\bibitem{albat} G.M. Viswanathan {\it et al.}, Nature {\bf 381}, 413
(1996).

\bibitem{Lev} P.  L\'evy, {\it  Th\'eorie de  l'addition des variables
al\'eatoires} (Gauthier-Villars, Paris, 1937).

\bibitem{CLT}  A.  Ara\'ujo  and  E.  Gin\'e,  {\it  The Central Limit
Theorem for Real and Banach Valued Random Variables} (Wiley, New York,
1980).

\bibitem{Proc} B.D. Hughes, M.F. Shlesinger, and E.W. Montroll, Proc.
Acad. Sci. USA {\bf 78}, 3287 (1981).

\bibitem{Compte} A. Compte, Phys. Rev. E {\bf 53}, 4191 (1996).

\bibitem{Foged} H.C. Fogedby, Phys. Rev. E {\bf 50}, 1657 (1993).

\bibitem{Tsa}  C.  Tsallis,  A.M.C.  Souza,  and  R.  Maynard, in {\it
L\'evy Flights and Related  Topics in Physics}, M.F.  Shlesinger, G.M.
Zaslavsky, and U.  Frisch, eds. (Springer, Berlin, 1995) p. 269.

\bibitem{MS} E.W. Montroll  and M.F. Shlesinger,  J. Stat. Phys.  {\bf
32}, 209 (1983).

\bibitem{AZ} P.A. Alemany and D.H. Zanette,  Phys. Rev. E {\bf 49},
956 (1994).

\bibitem{Ts1} C. Tsallis, J. Stat. Phys. {\bf 52}, 479 (1988).

\bibitem{Ts15} E.M.F. Curado and C. Tsallis, J. Phys. A {\bf 24}, L69
(1991) [corrigenda {\bf 24}, 3187 (1991); {\bf 25}, 1019 (1992)].

\bibitem{Ts2} C. Tsallis, Chaos, Solitons \& Fractals {\bf 6}, 539
(1995).

\bibitem{Ts3} C. Tsallis {\it et al.}, Phys. Rev. Lett. {\bf 75}, 3589
(1995).

\bibitem{ZA} D.H. Zanette and P.A. Alemany, Phys. Rev. Lett. {\bf 75},
366 (1995).

\bibitem{Wio} H.S. Wio, {\it  An Introduction to Stochastic  Processes
and Nonequilibrium Statistical Physics} (World Scientific,  Singapore,
1994).

\bibitem{LB}  A  formal  derivation  of  the  nonlinear  Fokker-Planck
equation  from  a  Langevin  equation  within  the  frame  of  Tsallis
statistics has recently been proposed in L. Borland, Phys. Rev. E {\bf
57}, 6634 (1998).


\bibitem{porous}  M.  Muskat,  {\it  The  Flow  of  Homogeneous Fluids
Through   Porous   Media}   (McGraw-Hill,   New   York,   1937);  P.Y.
Polubarinova-Kochina, {\it Theory of Ground Water Movement} (Princeton
University Press, Princeton, 1962).


\bibitem{surf} H. Spohn, J. Phys. (France) I {\bf 3}, 69 (1993).

\bibitem{films} J. Buckmaster, J. Fluid Mech. {\bf 81}, 735 (1977).

\bibitem{Mar} E.W. Larsen and G.C. Pomraning, SIAM J. Appl. Math.
{\bf 39}, 201 (1980).

\bibitem{others} W.L. Kath, Physica D {\bf 12}, 375 (1984).

\bibitem{scarce} J.D.  Murray, {\it  Mathematical Biology}  (Springer,
Berlin,  1989).

\bibitem{scarce2}  J.R.  King,  J.  Phys.  A {\bf 23}, 3681 (1990);
D.H.  Zanette, J. Phys. A {\bf 26}, 5339 (1993).

\bibitem{TsB} C. Tsallis and D.J. Bukman, Phys. Rev. E {\bf 54}, R2197
(1996).

\bibitem{Jou} A. Compte and D. Jou, J. Phys. A {\bf 29}, 4321 (1996).

\bibitem{Role} C.  Tsallis, R.S.  Mendes and  A.R. Plastino,  {\it The
role  of  constraints  within  generalized  nonextensive  statistics},
Physica A (1998, in press).
\end{references}
\end{document}